\documentclass[5p]{elsarticle}
\usepackage[usenames]{color}
\usepackage{epsfig}
\usepackage{amssymb}
\usepackage{amsmath}
\usepackage{multirow}
\usepackage{longtable}
\usepackage{xcolor}
\usepackage{ulem}
\usepackage{hyperref}
\hypersetup{
    pdfnewwindow=true,      
    colorlinks=true,       
    allcolors=SeaGreen
}

\newcommand\redsout{\bgroup\markoverwith{\textcolor{red}{\rule[0.5ex]{2pt}{0.4pt}}}\ULon}
\newcommand{\be}{\begin{equation}}
\newcommand{\ee}{\end{equation}}
\newcommand{\bc}{\begin{center}}
\newcommand{\ec}{\end{center}}
\newcommand{\beq}{\begin{eqnarray}}
\newcommand{\eeq}{\end{eqnarray}}
\newcommand{\er}{$\pm$}

\definecolor{grey}{rgb}{0.9,0.9,0.9}
\definecolor{black}{rgb}{0,0,0}

\biboptions{sort&compress}

\begin{document}
\begin{frontmatter}

\title{Note on the definitions of branching ratios of overlapping resonances}

\author[label3]{V.~Burkert}
\author[label4]{V. Crede}
\author[label3]{E.~Klempt}
\author[label1]{K.~V.~Nikonov}
\author[label8]{J.~A.~Oller}
\author[label9]{J.~R.~Pel\'aez}
\author[label9]{J.~Ruiz de Elvira}
\author[label1]{A.~Sarantsev}
\author[label7]{L.~Tiator}
\author[label1]{U.~Thoma}
\author[label6]{R. Workman}

\address[label3]{Thomas Jefferson National Accelerator Facility,
12000 Jefferson Avenue, Newport News, VA 23606, USA
}
\address[label4]{Department of Physics, Florida State University
Tallahassee, FL 32306-4350, U.S.A.
}
\address[label1]{Helmholtz-Institut f\"ur Strahlen- und Kernphysik der Universit\"at Bonn,
Nußallee 14-16, 53115 Bonn, Germany
}
\address[label7]{Institut f\"ur Kernphysik der Universit\"at Mainz,
Johann-Joachim-Becher-Weg 45, 55099 Mainz, Germany
}
\address[label6]{Institute for Nuclear Studies and Department of Physics,
The George Washington University, Washington, DC 20052, USA
}
\address[label8]{Departamento de F\'{\i}sica. Universidad de Murcia, 30071, Murcia, Spain.
}
\address[label9]{Departamento de F\'{\i}sica Te\'orica e IPARCOS. Facultad de Ciencias F\'{\i}sicas.  Universidad Complutense. Plaza de las Ciencias 1, 28040. Madrid, Spain
}

\date{\today}

\begin{abstract}
Branching ratios for the decay of hadrons with large width or near thresholds
depend on their definition.
We test different definitions and show that rather different branching ratios can be obtained.
For wide resonances and for sequential decays with wide intermediate resonances, integration
over the spectral functions is mandatory. The tests are performed exploiting the latest solution
of the Bonn-Gatchina multi-channel analysis and published values for residues of light scalar mesons.  For a resonance overlapping with a threshold, in case its pole lies in a non-adjacent sheet, we show how the total width, needed for the branching ratios, does not correspond to the imaginary part of the pole position. We use the Madrid-Krakow dispersive parameterizations to illustrate this situation with the $f_0(980)$.
\end{abstract}
\end{frontmatter}

\section{Introduction}
Quantum Chromodynamics at low energies features a large variety of resonances. Understanding their spectrum and their properties
is a topical aim. But what are the properties, and how are they defined? It is now consensus that the mass and the width of a
resonance should be identified with its pole position~(see~\cite{Hohler:1998inRPP} in Ref.~\cite{ParticleDataGroup:1998hll}). The
difference of mass determinations of the $\Delta(1232)$ resonance in Breit-Wigner parameterizations and pole extractions had
already been discussed in the 1974 edition of the Review of Particle Physics in ``Comments on the Mass and Width of
$\Delta(1232)$"~\cite{ParticleDataGroup:1974flk}. From then onwards, Breit-Wigner parameters and pole positions of baryons were
listed separately. The RPP-Baryon Group had decided in 2012 to move the pole properties of baryon resonances to the first place,
and the Breit-Wigner properties second, in order to emphasize the importance of the pole parameters. The change required some
time and was introduced in the 2018 edition~\cite{ParticleDataGroup:2018ovx}. For mesons, Breit-Wigner and pole parameters are
often still listed jointly~\cite{ParticleDataGroup:2022pjm}, leading to a larger spread of the results.

The intrinsic properties of resonances are reflected in
their coupling constants for decays into the different final states. The small ratio of the decay fractions of $f_2(1270)$ to
$K\bar K$ versus $\pi\pi$ and the large $K\bar K / \pi\pi$ ratio of $f_2(1525)$ identify the two mesons as mainly $(u\bar u+d\bar
d)/\sqrt2$ or $s\bar s$, respectively. The coupling constants of the light scalar mesons to two pseudoscalar mesons were used in~\cite{Oller:2003vf} to determine the mixing angle of the light scalar nonet. The significant contributions from cascade
decays via orbitally excited intermediate states of specific baryon resonances indicates an underlying three-quark dynamic
\cite{Thiel:2015kuc,CBELSATAPS:2015kka,CBELSATAPS:2022uad}. The sign of the imaginary part of the $K^-N\to\Lambda\pi$ and $\Sigma\pi$  amplitudes at the resonance
position reveals  their SU(3) structure~\cite{Burkert:2020inPDG}.

Instead of decay coupling constants, often only the decay branching ratios are given. Branching ratios can be determined from a
Breit-Wigner parameterization or at the pole position. In the PDG1986 edition~\cite{ParticleDataGroup:1986kuw}, branching ratios
were defined in the section on ``Kinematics" for elastic scattering and resonances represented by a Breit-Wigner function. In the
minireview on $N$ and $\Delta$ resonances in the Review of Particle Properties RPP2012~\cite{ParticleDataGroup:2012pjm}, the use
of residues to determine branching ratios and integration over the full spectral functions were suggested. In
RPP2016~\cite{ParticleDataGroup:2016lqr}, a new section on ``Resonances" was introduced~\cite{Asner:2016inPDG}, the scattering
amplitude and the definition of branching ratios were discussed in a much wider context and moved to this new section.
The notation used in our writing follows that used in the review on ``Resonances".

{
In this note, we discuss different prescriptions to define branching
ratios. None of
the prescriptions is strictly compatible with all requirements like analyticity, unitarity, and crossing symmetry of the amplitude. But to the best of our knowledge, no
such definition exists in the literature. 
The only definition in this paper fully consistent with
such requirements is the partial width of the $f_0(980)$ to 
two pions when obtained from a dispersive determination of its pole. But for the branching ratio, we would need the total width for which a model prescription is also needed.
Hence, we always need approximations for branching ratios.

For our discussion} we consider the dynamical part of the amplitude  for pion and photo-induced
scattering processes. If a resonance has spin, the decay products have -- in general -- non-flat angular distributions. The angular part
of the amplitude can be calculated in the helicity~\cite{Jacob:1959at} or in a spin-momentum operator-expansion formalism~\cite{Anisovich:2006bc}.

The note is organized as follows: after this short introduction and after defining some basic quantities, we give formulae used so far to calculate branching ratios. Within the
Bonn-Gatchina (BnGa) framework, fits to a large number of reactions are used to determine branching ratios for some low-mass resonances using different methods
for its determination.  First, for simple cases,
a conventional definition is used. Then, branching ratios are defined for overlapping resonances.
Wide resonances close to the threshold may require integration, for which expressions are given.
Next,  we recall formulae defining branching ratios from the residues of the resonance poles and show that these familiar definitions may need modifications when resonances are close to thresholds, as in the $f_0(980)$ 
case. We illustrate these modified definitions with a simple toy model providing a novel relation between residues
and branching ratios. We then apply this relation to the realistic case of the $f_0(980)$. 
We conclude with recommendations for the definition of branching ratios.

\section{Basic quantities}
The decay momentum $p_0$ for the decay of a particle $X$ with mass $M$ into two particles labeled 1 and 2 with masses $m_{i}$
in the rest frame of $X$ is given by
\begin{equation}
\hspace{-4mm} p_0 =\frac{\sqrt{\left(M^2-(m_1+m_2)^2)\,(M^2-(m_1-m_2)^2\right)}}{2\,M}\,.
\end{equation}
The invariant mass of $X$ can be calculated in any frame from the relation
\begin{equation}
M^2=s_0=(q_1+q_2)^2\,,
\end{equation}
where $q_i^\mu=(E_i,\vec p_i)$ are the four-momenta of the two particles.

The two-body phase space $\rho(s)$ is given by
\begin{equation}
\rho(s)=\frac{1}{16\pi}\frac{2p}{\sqrt{s}}\,,
\end{equation}
where $p=\sqrt{(s-(m_1+m_2)^2)\,(s-(m_1-m_2)^2)}/(2\,\sqrt s)$\,.

\section{Conventional branching ratios}
\subsection{\label{BW}Breit-Wigner resonances with constant widths}
The branching ratio of a narrow resonance far above decay thresholds, e.g. $J/\psi$, into a channel $a$, e.g. $\mu^+\mu^-$, can
be defined as number of  $J/\psi\to\mu^+\mu^-$ events overall $J/\psi$ events $N_{\rm tot}$

\begin{equation}\label{form4}
BR_a = N_a/N_{\rm tot} =  \Gamma_a/\Gamma_{\rm tot}\,,\qquad \sum_a\Gamma_a=\Gamma_{\rm tot}\,.
\end{equation}
$\Gamma_a$ is called the partial decay width and $\Gamma_{\rm tot}$ the total width of the resonance. $\Gamma_a$ can be determined by a formation experiment.
The cross-section is then written as
\begin{equation}
\sigma_{a\to b}(s) = \frac{\pi}{p^2}\frac{\Gamma_a\Gamma_b}{(M_{BW} - \sqrt{s})^2 +\frac14\,\Gamma^2_{\rm BW}}\,
\end{equation}
with $M_{\rm BW}$ and $\Gamma_{\rm BW}$ the Breit-Wigner mass and width, respectively. In the case of a $J/\psi$ resonance formed in $e^+e^-$ annihilation, the
right-hand side is multiplied by a factor ${2J+1}/ $ ${[(2s_1+1)(2s_2+1)]}=\frac34$ to account for the spin $J=1$ of the
resonance and $s_1=s_2=1/2$ of electron and positron.

The amplitude for $a\rightarrow b$ scattering is given by
\begin{equation}
{t}^{nr}_{ab}(s) =\frac12 \frac{\sqrt{\Gamma_a\Gamma_b}}{M - \sqrt{s} - i\frac12\,\Gamma_{\rm tot}}\,,
\end{equation}
or, alternatively, in a relativistic formalism
\begin{equation}
{t}_{ab}(s) = \frac{M\sqrt{\Gamma_a\Gamma_b}}{M^2 - s - i\,M\Gamma_{\rm tot}}\,.
\end{equation}

The partial decay width $\Gamma_a$ is used to define the coupling constants $g_a$
\begin{equation}
g_a^2 = 
\left|\,\frac{\sqrt{s_0}\,\Gamma_a}{\rho_a(s_0)}\,\right|\,. \label{nonrel}
\end{equation}
Note that for decays below threshold $(M<m_1+m_2)$, $\rho_a(s_0)$ can be a complex number.

\subsection{\label{relBW}Relativistic Breit-Wigner resonances
with energy-de\-pendent widths}

In more realistic cases, the resonance is not narrow and its mass is not far above the thresholds for relevant decays. In this
case, Eq.~(\ref{nonrel}) needs to be modified. With an angular momentum $l$ between the two outgoing particles, the amplitude is
suppressed by the angular-momentum barrier proportional to $(p/p_0)^l$ where the subscript ``0"  refers to the nominal mass
$M=\sqrt{s_0}$. At high momenta, the barrier becomes less important; a phenomenological Blatt-Weisskopf form-factor $F_l$ is
often applied that depends on the (running) breakup momentum and a range parameter $R$. We define the
orbital-angular-momentum-barrier factor for the resonance decay into channel $a$ by
\begin{equation}
n_a(s)= \left(\frac{p}{p_0}\right)^l \frac{F_l(p\,R)}{F_l(p_0\,R)}\,.
\end{equation}
Note that, to simplify the notation, the $a$ dependence of $p$ has been omitted. 
The  energy-dependent partial decay width is then given by
\begin{equation}
\Gamma_a(s) = \frac{g_a^2}{\sqrt{s_0}}\,\rho_a(s)\,n_a ^2(s)\,, \label{rel}
\end{equation}
where the coupling constants $g_a$ are now parameters of the scattering amplitude. With $z=(pR)^2$
\begin{eqnarray}
F_0^2= 1,\ \ & F_1^2=\frac{1}{1+z},\ \ &F_2^2=\frac{1}{9+3z+z^2}\,.
\end{eqnarray}
Higher order barrier factors can be found in Refs.~\cite{VonHippel:1972fg,Chung:1995dx}. The total width is given by the sum over
all contributing partial widths.

The associated scattering amplitude is written in the form
\begin{equation}
{t}_{ab}(s)=\frac{g_ag_b\sqrt{\rho_{a}(s)\rho_{b}(s)}\,n_a(s)n_b(s)}{M^2 - s - i\sum_i g_{i}^2 \rho_{i}(s)\,n_i ^2(s)}\,.
\label{amplitude}
\end{equation}

\subsection{\label{KM}Overlapping resonances}
In the case of overlapping resonances (with index $\alpha=1,2,\cdots$), the amplitudes $t_{ab}^\alpha$ cannot be added directly,
the resulting amplitude would violate unitarity. Often, the $K$-matrix formalism is applied,  which in its  classical form reads
\begin{equation}
    T(s)=K(s)\left[1-i\hat\rho(s) K(s)\right]^{-1},
    \label{eq:classicK}
\end{equation}
where  $(T)_{ab}=t_{ab}$ is the matrix amplitude, $\hat \rho$ is a diagonal matrix with elements $\rho_a$ and contributing resonances with index
$\alpha$ can be added in the $K$ matrix
\begin{equation}
K_{ab}(s) = \sum_\alpha \frac{g_a^{\alpha} g_b ^{\alpha}}{M_{\alpha}^2-s} + P_{ab}(s)\,,
\label{eq:Kmatrix-def}
\end{equation}
where the nonresonant ``background" is represented by a polynomial $P_{ab}(s)$. The formula (\ref{rel}) is still used to
calculate branching ratios. However, all quantities get an additional index $\alpha$ referring to the resonance $\alpha$.
$M_{\alpha}$ and $g^{\alpha}_{a,b}$  are the bare mass and bare couplings of the resonance $\alpha$, respectively. Exchanges in
the $t$-channel are represented by the exchange of Reggeons \cite{Denisenko:2016ugz}.

\subsection{\label{sequ}Sequential decays}
Sequential decays result in multi-body final states. We consider a concrete example: the decay of a
resonance with mass $M$ into a nucleon plus an intermediate resonance $r$, e.g. a
$\varrho$-meson that decays into $\pi\pi$. The final state is $N\pi\pi$, and the phase volume
$\rho_f(s)=\rho_3(s)$ is given by the $N\pi\pi$ phase volume
\begin{eqnarray}
\hspace{-4mm} \rho_3(s)=\!\!\int\limits_{(m_1+m_2)^2}^{(Re\sqrt{s}-m_3)^2} \!\!\frac{ds_{r}}{\pi}
\frac{\rho(s,s_{r},L_{r},R_{r})\,M_{r}\,\Gamma_{r}(s_{r})} {(M_{r}^2\!-\!s_{r})^2\!+\!(M_{r}\,\Gamma_{r}(s_{r}))^2}\,,
\label{rho3}
\end{eqnarray}
where $M_{r}$, $\Gamma_{r}$, $L_{r}$, and $R_{r}$ are the mass, width,
orbital angular momentum and
range parameter of the intermediate resonance. The latter decays
into two particles with momenta $k_1$ and $k_2$, $s_{r}=(k_1+k_2)^2$,
the relative momentum
\begin{equation}
\label{221219.1}
k=\frac{\sqrt{[s-(\sqrt{s_{r}}+m_3)^2][s-(\sqrt{s_{r}}-m_3)^2]}}{2\sqrt s}\,.
\end{equation}
The three-body phase space  in the quasi two-body approximation is given by
\begin{equation}
\rho(s,s_r,L_r,R_r)=\frac{1}{16\pi}\frac{2k}{\sqrt s}n_a(s)\,.
\end{equation}
For example, in the case of the  $\rho$-meson production, the width of the states can be defined via the P-wave two-pion phase
space and the corresponding decay coupling $g_{\pi\pi}$
\begin{equation}
M_\rho\Gamma_{\rm \rho}=g_{\pi\pi}^2\rho(M^2,M^2_{\rho},L_\rho=1,R_\rho),
\end{equation}
with an effective radius $R_\rho$ of the $\rho$-meson.

The function $\rho_3(s)$ has two threshold singularities in the complex plane of total energy squared $s$. The first one is the
cut on the physical axis starting at the squared sum of the final state particle masses, the second one is given by the cut in
the complex plane defined by the sum of masses of the spectator particle and the complex mass of the $\rho$-meson
$M_\rho-i\Gamma_\rho/2$.

\subsection{\label{intBW}Branching ratios by integration}
A broad resonance with a coupling to a final state, which is open only just above the Breit-Wigner mass, can decay into this final
state via the high-mass tail of the resonance.  In particular, the $N(1650)1/2^-$ has a Breit-Wigner mass of about 1650\,MeV, which falls below the
$K^+\Sigma^0$ threshold at 1687 MeV. The branching ratio defined in section~\ref{relBW},  or later in \ref{f-BW}, yields zero, which is
counter-intuitive since there could be some $N(1650)1/2^-$ $\to K^+\Sigma^0$ (or $K^0\Sigma^+$) decays. A physically more
intuitive definition integrates the cross section
\begin{eqnarray}
\hspace{-7mm} BR_a =  \!\!\int\limits_{\rm threshold}^{\infty}\!\!\frac {ds}{\pi} \frac{ g_a^2\rho_{a}(s) n_a ^2(s)}{(M^2-s)^2+
(\sum\limits_i g_{i}^2\rho_{i}(s) n_i ^2 (s))^2}, \label{br4}
\end{eqnarray}
and the $N(1650)1/2^-$ $\to K\,\Sigma$ decay branching ratio can be non-zero.

\section{Branching ratios from residues}
\label{residues}
Residues of transition amplitudes can be calculated through a contour integral of the amplitude $T_{ab}$ around the pole position
$W_p=\sqrt{s_p}=M_p-i\Gamma_p/2$ in the energy plane\footnote{Note that in Eq.~\eqref{contour}, we use for the residue an unsual convention, since the integral is over $\sqrt s$ instead of $s$.}
\begin{equation}
Res(a\to b)=-\oint\frac{d\sqrt s}{2\pi i}\,\,t_{ab}(s). \label{contour}
\end{equation}
For elastic scattering of channel $a \to a$, e.g. for $\pi N\to \pi N$, this gives a good approximation for the elastic residue
\begin{equation}
Res(a\to a)=\frac{g_a^2}{2W_{p}}\rho_a(W_p^2)n_a(W_p^2),
\end{equation}
and the $\pi N$ partial width and consequently the branching ratios at the pole position
\begin{equation}
\Gamma^{\rm pole}_{\pi N} = 2\,|Res (\pi N\to \pi N)|\,.
\end{equation}
For the calculation of any partial decay width, $a \to b$, we use the factorization
\begin{equation}
Res(a\to b)^2 = Res(a\to a) Res(b\to b),
\end{equation}
and can calculate the branching ratio from residues in further good approximation ($W_p \to M$) with
\begin{equation}
\Gamma^{\rm pole}_{b}= 2\,|Res (b\to b)| =  \frac{g_b^2}{M}\rho_b(M^2) \,. \label{br5}
\end{equation}

\section{Fits and fit results  in BnGa approach }
\subsection{PWA method}
The partial-wave analysis performed within the BnGa formalism is described in detail in a number of
publications~\cite{Anisovich:2006bc,Denisenko:2016ugz,Anisovich:2004zz,Anisovich:2007zz}. Here, we give a short outline.

The transition amplitude $\hat T$ contains poles due to resonances and background terms and,  in the BnGa fits it is 
described by a modified K-matrix
 \begin{equation}
 \mathbf{\hat T}(s) = \mathbf{\hat K}(s)\;\left(\mathbf{\hat I}\;-\mathbf{\hat B(s) \hat K(s)}\right)^{-1}\,, \label{k_matrix}
\end{equation}
where $\mathbf{\hat K}$ is a matrix whose elements have the functional form of Eq.~\eqref{eq:Kmatrix-def}, and $\mathbf{\hat B}$ is a diagonal matrix of loop diagrams with an imaginary part equal to the corresponding phase space volume

\begin{equation}
\hat B_a(s)= Re B_a(s)+\;i\rho_a(s)\,.
\end{equation}
If the real part of the loop diagram is neglected, this method corresponds to the classical K-matrix approach, 
Eq.~\eqref{eq:classicK}. In the fits one
subtraction is used to calculate the elements of the B-matrix
\begin{eqnarray}
\hspace{-5mm}B_{a}(s)&=&b_i+(s-(m_{1a}+m_{2a})^2)\times\\
&&\hspace{-4mm}\int\limits_{(m_{1a}+m_{2a})^2}^\infty\!\!\!\! \frac{ds'}{\pi}
\frac{\rho_a(s',L,r)}{(s'-s-i\epsilon)(s'-(m_{1a}+m_{2a})^2)}\,,\nonumber
\end{eqnarray}
where $\epsilon$ goes to zero. The $b_a$ are subtraction constants, determined by the fit, and $m_{1a}, m_{2b}$ are masses of the particles in channel $a$.

\subsection{\label{data}The database}
The BnGa  group uses a large body of pion and photo-induced reactions including data
from Bonn, JLab and Mainz. These data
sets are listed on the BnGa web page (pwa.hiskp.uni-bonn.de/).

\subsection{\label{respol}Poles and residues}
The low-energy region is fitted with poles describing $\Delta(1232)3/2^+$,
$N(1440)1/2^+$,  $N(1520)3/2^-$, $N(1535)1/2^-$, $\Delta(1600)1/2^-$,
$N(1650)1/2^-$, and poles at higher masses.  We \,discuss \,here \,the \,two
\,resonances \,$N(1535)1/2^-$ \,and $N(1650)1/2^-$~to~demonstrate the effect
of overlapping resonances, and  $N(1520)3/2^-$ to study the model dependence
of the $N(1520)3/2^-\to N\rho$ branching ratio.

The amplitude was searched for poles in the complex $s$-plane. The pole positions
were identified and are listed in Table~\ref{poles}. The Breit-Wigner masses and widths
were determined with the method described in Section~\ref{f-BW} below.

The residues were calculated by integration along a closed contour around the pole 
(see Eq.~(\ref{contour})). The numerical
values are presented in Table~\ref{res}.

\subsection{\label{f-BW}Breit-Wigner mass and width}
\begin{table}[pt]
\caption{\label{poles}Pole position and Breit-Wigner mass and width of
low-mass baryon resonances. Masses and widths are given in GeV.}
\renewcommand{\arraystretch}{1.15}
\bc
\begin{tabular}{cccc}
\hline\hline
Resonance & $W_{pole}$ & $M_{\rm BW}$ & $\Gamma_{\rm BW}$\\
\hline
$N(1535)1/2^-$                &$1.495 - i0.063$       & 1.517       &        0.124\\
$N(1650)1/2^-$                &$1.645 - i0.049$       & 1.650       &       0.099\\
$N(1520)3/2^-$               &$1.506 - i0.056$       & 1.515       &        0.113\\
\hline\hline
\end{tabular}
\vspace{-2mm}\ec
 \caption{\label{res}Residues (in MeV) for $\pi N\to \rm resonance \to final ~ state$ transitions
 from the BnGa PWA.
}
\renewcommand{\arraystretch}{1.15}
\bc
{
\begin{tabular}{lcccc}
\hline\hline
$N(1535)1/2^-$$\to$&\hspace{-4mm}$\pi N$  &  $\eta N$ & $ K^+ \Lambda$ & $K\,\Sigma$\\\hline
 modulus          &\hspace{-4mm}29\er4&31\er4&45\er8   &\\
 phase ($^\circ$) &\hspace{-4mm}$-33$\er7 &$-80$\er8&$-135$\er18 &\\
\hline
$N(1650)1/2^-$$\to$&\hspace{-4mm} $\pi N$  &  $\eta N$ & $ K^+ \Lambda$ & $K\,\Sigma$\\\hline
 modulus          &\hspace{-4mm}24\er5&$-22$\er5     &9\er3& 51\er12\\
 phase ($^\circ$) &\hspace{-4mm}$-58$\er15 &$-16$\er14 &75\er20 &60\er20\\
\hline
$N(1520)3/2^-$$\to$&\hspace{-4mm} $\pi N$  &  $\pi\Delta(S)$ & $\pi\Delta(D)$ & $\rho N(\frac 32 S)$ \\\hline
 modulus          &\hspace{-4mm}35\er4   &$-16$\er4     &$-12$\er5&$-14$\er4\\
 phase ($^\circ$) &\hspace{-4mm}$-16$\er4  &$-2$\er12 &$-45$\er18 &$-40$\er15\\
\hline\hline
\end{tabular}
}\ec
\end{table}

Even for a simple relativistic Breit-Wigner amplitude (only one decay channel with an energy-dependent width), the Breit-Wigner mass and width do not coincide with the position of
the pole in the complex $W$ plane. In order to determine Breit-Wigner parameters of a resonance, the BnGa collaboration fitted a
Breit-Wigner amplitude so that its pole matches the position of the pole position of the full $K$-matrix fit. In this way, a
Breit-Wigner amplitude is constructed that has the correct pole position and is free from any ``background" or influence of
neighboring resonances. In the BnGa approach, the amplitude is approximated by
\begin{equation}
\hspace{-2mm} t_{ab}^{\alpha}(s)=\pm\frac{f\,g_{a}^{\alpha}g_{b}^{\alpha} \sqrt{\rho_{a}^{\alpha}(s)\rho_{b}^{\alpha}(s) }\
n_a^{\alpha}(s)n_b^{\alpha}(s)}{(M^{\alpha})^2-s -if\sum\limits_i (g_i^{\alpha})^2\rho_i^{\alpha}(s) (n_i^{\alpha}(s))^2}\,,
\end{equation}
where the sign depends on the phase difference between $g_{a}^{\alpha}$ and $g_{b}^{\alpha}$. The parameters $M^\alpha$ and $f$
were calculated to reproduce the pole position of the amplitude.
 In typical situations, $f$ adopts values of about 0.85 to 1.2. Note that residues calculated
from this amplitude are shifted with respect to the K-matrix values. In other approaches like MAID, SAID, or our toy model, the Breit-Wigner parameters are fitted
directly, and the pole position is deduced from the amplitude. For such approaches, $f=1$.

\begin{table*}
\caption{\label{results}Decay branching ratios of $N(1535)1/2^-$, $N(1650)1/2^-$ and
 $N(1520)3/2^-$ using different definitions: ``nr-BW" (Section\,\ref{BW}), ``rel-BW" (Section\,\ref{relBW}),
 ``K-matrix" (Section\,\ref{f-BW}), ``residue" (Section\,\ref{residues}), ``integration (Section\,\ref{intBW})".
The statistical errors are very small, hence no errors are given. The sum of branching ratios derived from residues and from
integration deviates from unity. Normalized values are given in parentheses. } \vspace{2mm}\renewcommand{\arraystretch}{1.4}
\begin{center}
{
\begin{tabular}{lcccrrrr}
\hline\hline & nr-BW & rel-BW & $K$-matrix & \multicolumn{2}{c}{residue}& \multicolumn{2}{c}{integration} \\
\hline
$N(1535)\to\pi N$             & 63\% & 48\%  & 49\%  &46 &\hspace{-4mm}(59)\% & 57 &\hspace{-4mm}(48)\%  \\
$N(1535)\to\eta N$           & 22\% & 38\%  & 37\%  &16 &\hspace{-4mm}(21)\% & 22 &\hspace{-4mm}(19)\%  \\
$N(1535)\to K^+\Lambda$&  0\% &  0\%   &  0\%   &0 &\hspace{-4mm}(0)\%  & 8  &\hspace{-4mm}(7)\%    \\
$N(1535)\to K \Sigma$      &  0\% &  0\%   &  0\%   &0 &\hspace{-4mm}(0)\%  &  12 &\hspace{-4mm}(10)\%   \\
$N(1535)\to \pi\Delta$      &  3\% &  3\%   &  3\%   &4 &\hspace{-4mm}(5)\%  &    4 &\hspace{-4mm}(3)\%     \\
$N(1535)\to \rho N$(S)     &  2\% &  2\%   &  2\%   &3 &\hspace{-4mm}(4)\%  &    3 &\hspace{-4mm}(3)\%     \\
$N(1535)\to \rho N$(D)    & 0.2\%&  0.2\% &  0.2\%&  0.2&\hspace{-4mm}(0.3)\%& 1.2 &\hspace{-4mm}(1)\%   \\
$N(1535)\to \sigma N$     & 4\%  &  4\%    &  4\%   &4 &\hspace{-4mm}(5)\%  & 5    &\hspace{-4mm}(4)\%    \\
$N(1535)\to \pi N(1440)$  & 5\%  &  5\%    &  5\%   &5 &\hspace{-4mm}(6)\%  & 6   &\hspace{-4mm}(5)\%     \\
$\sum$                          & 100\%&100\%  &100\% & 78 & \hspace{-4mm}(100)\% & 118 & \hspace{-4mm}(100)\%\\
\hline \hline
$N(1650)\to\pi N$             & 42\%   & 42\%  & 42\%  &48&\hspace{-4mm}(42)\% & 34&\hspace{-4mm}(34)\%    \\
$N(1650)\to\eta N$           & 32\%   & 32\%  & 32\%   &37&\hspace{-4mm}(32)\% & 23&\hspace{-4mm}(23)\%    \\
$N(1650)\to K^+\Lambda$& 2\%    &  2\%  &  2\%         &2&\hspace{-4mm}(2)\% &  1 &\hspace{-4mm}(1)\%     \\
$N(1650)\to K\Sigma$       &  0\%   & 0\%   &  0\%     &0&\hspace{-4mm}(0)\% & 18&\hspace{-4mm}(18)\%     \\
$N(1650)\to \pi\Delta$      &  4\%   &  4\%   &  3\%   &4&\hspace{-4mm}(3)\% & 3  &\hspace{-4mm}(3)\%     \\
$N(1650)\to \rho N$(S)     &  7\%   &  7\%   &  7\%    &8&\hspace{-4mm}(7)\% & 7  &\hspace{-4mm}(7)\%     \\
$N(1650)\to \rho N$(D)    &  4\%   &  4.5\% &  4\%     &5&\hspace{-4mm}(4)\% & 6  &\hspace{-4mm}(6)\%     \\
$N(1650)\to \sigma N$     & 5\%    &  5.5\% &  5\%     &6&\hspace{-4mm}(5)\% & 5  &\hspace{-4mm}(5)\%     \\
$N(1650)\to \pi N(1440)$  & 4\%    &  4\%   &  4\%     &5&\hspace{-4mm}(4)\%  & 4   &\hspace{-4mm}(4)\%   \\
$\sum$                       & 100\%  & 100\% &100\%   &115&\hspace{-4mm}(100)\% &101&\hspace{-4mm}(100)\%   \\
\hline \hline
$N(1520)\to\pi N$                  & 62\%  & 57\%  & 58\%  & 60&\hspace{-4mm}(58)\%  & 49&\hspace{-4mm}(45)\%     \\
$N(1520)\to \pi\Delta(S)$        & 13\%  & 14\%  & 14\%    & 13&\hspace{-4mm}(13)\%  & 13&\hspace{-4mm}(12)\%      \\
$N(1520)\to \pi\Delta(D)$        & 8\%   & 9.5\% &  9\%    & 8&\hspace{-4mm}(8)\%    & 14&\hspace{-4mm}(13)\%       \\
$N(1520)\to \rho N(\frac 32 S)$& 10\% & 11.5\%& 11\%       & 10&\hspace{-4mm}(10)\%  & 24&\hspace{-4mm}(22)\%        \\
$N(1520)\to \rho N(\frac 12 D)$& 0.1\%& 0.1\% & 0.1\%      & 0.1&\hspace{-4mm}(0.1)\% & 0.3&\hspace{-4mm}(0.3)\%      \\
$N(1520)\to \rho N(\frac 32 D)$& 0.0\%& 0.0\% & 0.0\%      & 0&\hspace{-4mm}(0)\%    & 0.2&\hspace{-4mm}(0.2)\%     \\
$N(1520)\to \sigma N$            & 7.0\% & 7.2\% & 7.3\%   & 7&\hspace{-4mm}(7)\%    & 11&\hspace{-4mm}(10)\%        \\
$\sum$                              & 100\% &100\% &100\%   &98&\hspace{-4mm}(100)\% & 112&\hspace{-4mm}(100)\%   \\
\hline\hline
\end{tabular}
}\end{center}
\end{table*}

\subsection{\label{br}Branching ratios}
For the fits with different definitions for the branching ratios we replaced the three poles by non-relativistic or relativistic
Breit-Wigner amplitudes. The fits became worse: the addition of two Breit-Wigner amplitudes violates unitarity, and the data are
no longer well described.  The other branching ratios are all derived from our best fit.

The results on branching ratios using this definition are given in Table~\ref{results}. In the case of overlapping resonances,
the fit with relativistic Breit-Wigner amplitudes is still inappropriate to describe the $\pi N$ elastic scattering amplitudes.

\section{\boldmath Branching ratios from poles in non-adjacent     sheets, the $f_0(980)$ case}

So far, we have been dealing mostly with relatively well-behaved cases: (i) poles associated with resonances that appear in the
Riemann sheet contiguous to the physical one once all channels of interest are open, and (ii) poles well isolated from 
other singularity structures. However, for resonances lying close to the threshold, poles in other sheets may
become relevant and then the discussion requires some modification.  This is not a purely academic discussion; it is indeed the
case of the $f_0(980)$ scalar meson, which we will use as an example. The same situation can also occur for the $1/2^-$ baryon
resonances $N(1535)$, close to the $\eta N$ threshold, $N(1650)$, close to the $K \Sigma$ threshold and $N(1895)$, close to the
$\eta' N$ threshold.

The $f_0(980)$ pole parameters were determined  from a dispersive data analysis \cite{Garcia-Martin:2011iqs} constrained to
satisfy forward dispersion relations, sum rules, and two kinds of Roy-like equations. The latter provide a model-independent
continuation to the complex plane \cite{Garcia-Martin:2011nna}, resulting in an $f_0(980)$ pole at
$(996\pm7-i\,25^{+10}_{-6})\,$MeV, with a $\vert g_{\pi\pi}\vert=2.3\pm0.2\,$GeV coupling to two pions. One is then tempted, as usual, to identify the total width as minus twice the imaginary part of the pole position and the partial width to two pions
defined in Eq.~\eqref{rel}, $\Gamma_{\pi\pi}=\vert g_{\pi\pi}\vert^2 \rho_{\pi\pi}(M^2)/M$. However this would lead to $\Gamma_{pole}=50^{+20}_{-12}\,$MeV and $\Gamma_{\pi\pi}=100^{+20}_{-17}\,$MeV. The partial width would become {\it twice} the  pole width!  Such a naive interpretation is wrong.
This issue was already noticed in~\cite{Wang:2022vga}, where the relation between the pole width and residues in a two-channel approximation was established (Eq.~\eqref{eq:GIInottot} below).

To explain this apparent contradiction, let us first recall that no four-pion states are observed in $\pi\pi$ scattering below 1
GeV, for all means and purposes the first inelastic threshold for the $\pi\pi$ $S0$ wave is that of $K\bar K$ \footnote{As customarily done, we consider the isospin limit where there is only one $K\bar K$ threshold, instead of
two, neutral and charged thresholds separated by 8 MeV.}.
Then, the problem is that the $f_0(980)$ pole in \cite{Garcia-Martin:2011nna} is not in the contiguous sheet above both the two-pion and the $K \bar K$ thresholds (third sheet), where
all the considerations made in previous sections would apply. Actually, although its nominal mass $996\pm7\,$MeV is above the $K\bar K$
threshold, its associated pole lies in the second sheet, which is only contiguous to the physical sheet {\it
below} the  $K \bar K$ threshold. 
In such a case, the familiar identification of the total width with half the imaginary part of the pole position no longer holds. This is because, in this unusual sheet, there is a change of sign in the kaon phase-space contribution with respect to the familiar situation. This was first pointed out in~\cite{Wang:2022vga} but we will illustrate it next with a very simplified toy model.

\subsection{Toy model}

The aim of this subsection is to illustrate, in the simplest possible setting, the relation between total or partial widths and pole parameters when the pole under study has a mass slightly above a threshold but lies in the contiguous sheet below that threshold. It does not address any other features like other backgrounds or nearby resonances. Therefore it does not apply directly to the $f_0(980)$ case,
whose realistic amplitude will be discussed in the following subsection.

Hence, we will construct a toy model of a resonant $S$-wave scattering with two-channels $a=1,2$ (we could think of $1=\pi\pi$  and $2=K\bar K$\,). Let us then recall that for an $S$-wave, $n_a(s)=1$ and $\rho_a(s)=\sqrt{s-s_{a}}/(16\pi\sqrt{s})$, with $s_a =s_{th,\,a}$ the first and second thresholds.  Then,  
following Eq.~(\ref{amplitude}) we can write the $t_{11}$ matrix-element in 
the upper half of the physical or first sheet, $Im\, s\geq 0$,  
as follows
\begin{equation}\label{RS-I}
 t_{11}^{I}(s) = \frac{g_1^2\, \rho_1(s)}{M^2-s-i\,(g_1^2\, \rho_1(s) + g_2^2\, \rho_2(s))}\,.
\end{equation}

The analytic continuation through the real axis leads us to the adjacent sheet in the lower half plane ($Im\, s<0$).\footnote{The cut of $\sqrt{s-s_a}$ appearing in the definition of $\rho_a$ is chosen to lie on $s\le s_a$ in the real axis.} However, this sheet is different depending on where the real axis is crossed.
In particular, the adjacent sheet
 between the first and second thresholds, $s_1\leq s\leq s_2$, is usually called the second sheet.
Here, the amplitude adopts the form
\begin{equation}\label{RS-II}
t_{11}^{II}(s) = \frac{g_1^2\, \rho_1(s)}{M^2-s-i\,(g_1^2\, \rho_1(s) - g_2^2\, \rho_2(s))}\,.
\end{equation}
Note that by crossing the real axis from the first to the second sheet the sign in front of $\rho_2(s)$ has changed, whereas that of 
$\rho_1(s)$ stays the same.

Above the $K\bar K$ threshold  in the real axis ($s>s_2$),
the adjacent sheet is now called the
third Riemann sheet. Expressing the amplitude with the standard square-root functions, the formula is identical to Eq.~(\ref{RS-I}) of the first sheet
\begin{equation}\label{RS-III}
t_{11}^{III}(s) = \frac{g_1^2\, \rho_1(s)}{M^2-s-i\,(g_1^2\, \rho_1(s) + g_2^2\, \rho_2(s))}\,,
\end{equation}
but now it applies in the lower half plane.





Let us now consider the partial wave near $s=M^2$, in the scenario when it is dominated by an isolated pole in the lower plane of
the third sheet, located at $s_{R,III}$. If the partial widths are defined as $\Gamma_a\equiv g_a^2\rho_a(M^2)/M>0$, then, in the
$\rho_a(s)\simeq\rho_a(M^2)$ and $\Gamma_{\rm tot}^2/M^2\ll 1$ approximation, with Eq.~\eqref{RS-III} we find
\begin{eqnarray}
t^{III} _{11}(s)&\simeq&\frac{g_1^2\,\rho_1(s)}{s_{R,III}-s}\\
&\simeq&\frac{g_1^2\,\rho_1(s)}{M^2- i\,(g_{1}^2\,\rho_{1}(s)+g_{2}^2\,\rho_{2}(s))-s}\nonumber\\
&\simeq& \frac{g_1^2\,\rho_1(s)}{(M-i(\Gamma_1+\Gamma_2)/2)^2-s}\nonumber\\
&=&\frac{g_1^2\,\rho_1(s)}{(M-i\Gamma_{\rm tot}/2)^2-s}.\nonumber \label{eq:twithGammas}
\end{eqnarray}
From the above follows the usual relation
 \begin{equation}
     \sqrt{s_{R,III}}\simeq M-i\Gamma_{\rm tot}/2=M-i(\Gamma_1+\Gamma_2)/2,
     \label{eq:poleparam}
 \end{equation}
between the pole position and resonance parameters. Obviously, the branching ratios satisfy  $BR_a=\Gamma_a/\Gamma_{\rm tot}<1$.

In contrast, imagine that there was a pole not in the adjacent sheet, but in the second one, located at $s_{R,II}$. If this pole
is well above $s_2$ it will barely influence what we see for physical real values of $s$. Nevertheless, if it lies
above but very close to $s_2$, it can largely influence what we see in the first sheet in the real axis below $s_2$. In addition, it can still exert some influence in the real axis above that threshold.
With Eq.~\eqref{RS-II} we find:
\begin{eqnarray}
 t^{II} _{11}(s)
&\simeq&\frac{g_1^2\,\rho_1(s)}{s_{R,II}-s}\\
&\simeq&\frac{g_1^2\,\rho_1(s)}{M^2- i\,(g_{1}^2\,\rho_{1}(s)-g_{2}^2\,\rho_{2}(s))-s}\nonumber\\
&\simeq&
 \frac{g_1^2\,\rho_1(s)}{(M-i(\Gamma_1-\Gamma_2)/2)^2-s}\nonumber\\
 &=& \frac{g_1^2\,\rho_1(s)}{(M-i\Gamma_{II}/2)^2-s}.\nonumber
\label{eq:tsecond}
\end{eqnarray}
Hence, for poles in the second sheet in the above circumstances, the familiar relationship between the pole position and
resonance  parameters given in Eq.~\eqref{eq:poleparam} does not apply. Instead, we find
\begin{align} \label{eq:GIInottot}
\sqrt{s_{R,II}}\,\simeq\, & M-i \Gamma_{II}/2,\\
\Gamma_{II}\,\equiv\, & \Gamma_1-\Gamma_2.\nonumber
\end{align}
Therefore, 
\begin{equation}
\Gamma_{II} \neq  \Gamma_{\rm tot}.
\end{equation}
Now, the branching ratios cannot be  $\Gamma_a/\Gamma_{II}$, since then we would find $\Gamma_1>\Gamma_{II}$ and $BR_1>1$, which
is absurd. Nonetheless, attending to the calculation of cross sections influenced by the exchange of such resonance, it is still
appropriate to define $BR_a=\Gamma_a/\Gamma_{\rm tot}$ as well.

This simple model does not directly apply to the $f_0(980)$ since important features, like the presence of the very broad $f_0(500)$ and $f_0(1370)$ resonances, as well as possible $4\pi$ contributions, have been neglected. As a consequence, it does not even get close to describing the scattering data.
Nevertheless, the relation between the pole parameters and the total and partial widths in Eq.~\eqref{eq:GIInottot} still holds within the fairly reasonable two-channel approximation.

\subsection{The $f_0(980)$ branching ratios from its pole}
The dispersive analysis of $\pi\pi\to\pi\pi$ scattering data in~\cite{Garcia-Martin:2011nna} finds a $f_0(980)$ pole that lies in the second sheet. Then, and in view of the previous pedagogical model, it is clear that $\Gamma_{pole}=\Gamma_{II}=50^{+20}_{-12}\,$MeV does not correspond to the total width $\Gamma_{\rm tot}$. As
explained above, the partial width to $K\bar K$ contributes negatively. Nevertheless, the partial width
\begin{equation}
\Gamma_{\pi\pi}=\frac{g_{\pi\pi}^2}{M}\rho_{\pi\pi}(M^2)=100^{+20}_{-17}\, {\rm MeV},
\end{equation}
is still correct.

Now, in a two-channel approximation, i.e., considering just the $1=\pi\pi$ and $2=K\bar K$ channels, we can use again Eq.~\eqref{eq:GIInottot} above. Including the correlations from the Madrid-Krakow dispersive calculation, we thus obtain,
sequentially,
\begin{eqnarray}
&&\Gamma_{K\bar K}\simeq\Gamma_{\pi\pi}-\Gamma_{II}=50^{+26}_{-21}\, {\rm MeV}, \label{r}\\
&&\Gamma_{\text{tot}}\simeq\Gamma_{\pi\pi}+\Gamma_{K\bar K}= 151^{+44}_{-37}\,{\rm MeV} ,\nonumber\\
&&{\rm BR}_{\pi\pi}\simeq0.67\pm0.07,\nonumber\\
&&{\rm BR}_{K\bar K}\simeq0.33\pm0.07,\nonumber\\
&& r_{K\bar K/\pi\pi}={\rm BR}_{K\bar K}/{\rm BR}_{\pi\pi}\simeq0.49\pm0.11,\nonumber
\end{eqnarray}
where the approximate symbol is due to our neglecting other channels than $\pi\pi$ and $K\bar K$. These results are very
consistent with the list of seven values provided in the Review of Particle Physics, whose central values for
$\Gamma_{\pi\pi}/(\Gamma_{\pi\pi}+\Gamma_{K\bar K})$ range from 0.52 to 0.84. In particular, it is remarkably consistent with the
value of $0.75\pm 0.20$ obtained from the chiral unitary approach in~\cite{Wang:2022vga,Guo:2012yt,Guo:2012ym}. Finally, in case the reader is interested in a
parameterization and not just the numerical outputs of the dispersive integrals, almost identical values can be obtained with the
``global" $\pi\pi\to \pi\pi$ description~\cite{Pelaez:2019eqa,Pelaez:2022qby} of  the dispersive output of the Madrid-Krakow
that makes use of Chew-Mandelstam
functions similar to the $B(s)$ function used in~\eqref{k_matrix} above, as well as a conformal expansion that includes other
backgrounds like those of the $f_0(500)$ or even additional $4\pi$ inelasticities, all combined to respect unitarity.

As a final step, we use the integration method, Eq.~(\ref{br4}), to determine the branching ratios of $f_0(980)$ to $K\bar K$ and
$\pi\pi$. We use the scalar isoscalar amplitude derived in a fit data on $J/\psi$ radiative decays and other data. The $f_0(980)$
pole was found at $(1014\pm 8 -i\, 35.5\pm 5)$\,MeV~\cite{Sarantsev:2021ein}. The $f_0(980)$ width is considerably  larger than the
width given above but fully compatible with the range reported by the Particle Data Group of 10 to
100\,MeV~\cite{ParticleDataGroup:2022pjm}. For the ratio we obtain
\begin{equation}
 r_{K\bar K/\pi\pi}=0.40\pm0.07,
\end{equation}
in good agreement with~\eqref{r}.

\section{Conclusions}
We have studied different definitions of branching ratios. For baryons, there is a considerable spread of results when different
definitions are used. Fits exploiting Breit-Wigner amplitudes or a $K$-matrix yield branching ratios that sum up to 100\% when
all permitted decay modes are included. This normalization does not hold true when branching ratios are integrated over the
available phase space or when they are defined via the transition residues of the complex amplitudes. Up to 20\% normalization
errors are observed.

Large differences are seen when the different definitions of branching ratios are used. The $N(1535)1/2^-\to N\eta$ branching
ratio is found with a 21\% (normalized) branching ratio when calculated from the $N\eta\to N\eta$ residue, 19\% after integration
over the phase space, and 38\% when two  relativistic Breit-Wigner amplitudes are added. The corresponding numbers for the
$N(1650)1/2^-\to N\eta$ branching ratios are 32\%, 23\%, and 32\%. The $N\eta$ threshold hits the $N(1535)1/2^-$ mass
distribution but is far below the $N(1650)1/2^-$ mass. In the former case, the branching ratios are much more sensitive to their
precise definition.

A second example is the $N(1520)3/2^-\to N\rho$ decay. The HADES Collaboration used the $K$-matrix approach and integrated over
the $\rho$ spectral function \cite{HADES:2020kce} as outlined in section~\ref{sequ}. Their result of (12.2\er 1.2)\% is
compatible with our finding. In 1992, Manley and Saleski reported (21\er4)\% \cite{Manley:1992yb} and had integrated over both,
the intermediate $\rho$ resonance and the $N(1520)3/2^-$ line shape. We confirm their findings.

What definition should be used? For narrow resonances far above a threshold, any definition can be
used. In other situations, branching ratios are best represented by the numbers obtained after integration over the
mass distribution of the produced resonance. In the case of sequential decays, a double integration may be needed, taken into
account the mass distribution of the primarily produced resonance and the spectral function of the intermediate resonance. These numbers
correspond to the expectation we have for an isolated resonance in an ideal world with no background and no interference with
other resonances: How often do we observe a final state when the resonance decays? If we want to test symmetry relations, e.g.
SU(3) symmetry, the coupling constants should be used directly.

We have also provided a new definition of branching ratios for resonances that overlap with nearby thresholds and whose
pole lies in a non-adjacent sheet. In this case, the usual relation between the imaginary part of the pole position and the total
width does not hold \cite{Wang:2022vga}, but branching ratios can still be defined from the pole and its residues, for which we
provide a prescription. The results for the $f_0(980)$ branching ratios, using the pole residues and positions from the
$\pi\pi\to \pi\pi$ dispersive analysis of \cite{Garcia-Martin:2011nna}, are in good agreement with the present status in the Review of Particle Physics.

\section*{Acknowledgments}
This work has been supported in part by the NSFC and the Deutsche Forschungsgemeinschaft (DFG, German Research Foundation) through the funds
provided to the Sino-German Collaborative Research Center TRR110 “Symmetries and the Emergence of Structure in QCD”
(NSFC Grant No. 12070131001, DFG Project-ID 196253076 - TRR 110), by the programme ”Netzwerke 2021”, an initiative of the Ministry of Culture and Science of the State of Northrhine Westphalia (project ”NRW-FAIR”, ID: NW21-024-C), by the U.~S.~Department of Energy, Office of Science, Office of Nuclear Physics under Award No. DE--SC--0016582.
The work by JRP and JRE is supported in part by the  Ministerio de Ciencia e Innovación grant PID2019-106080GB-C21.
JRE also acknowledges financial support from the Ramón y Cajal program (RYC2019-027605-I) of the Spanish MINECO. The work of JAO was supported in part by the MICINN AEI (Spain) Grant No. PID2019–106080GB-C22/AEI/10.13039/501100011033.
The work by JAO, JRP, JRE, and UT is also partially supported by the EU Horizon 2020 research and innovation programme, STRONG-2020 project, under grant agreement No. 824093.


\begin{thebibliography}{99}
\bibitem{Hohler:1998inRPP}
G.~H\"ohler,
``Against Breit-Wigner parameters -- a pole-emic",
in Ref.~\cite{ParticleDataGroup:1998hll}

\bibitem{ParticleDataGroup:1998hll}
C.~Caso \textit{et al.} [Particle Data Group],
``Review of particle physics. Particle Data Group,''
Eur. Phys. J. C \textbf{3}, 1-794 (1998).

\bibitem{ParticleDataGroup:1974flk}
V.~Chaloupka \textit{et al.} [Particle Data Group],
``Review of particle physics. Particle Data Group,''
Phys. Lett. B \textbf{50}, 1-198 (1974)

\bibitem{ParticleDataGroup:2018ovx}
M.~Tanabashi \textit{et al.} [Particle Data Group],
``Review of Particle Physics,''
Phys. Rev. D \textbf{98}, no.3, 030001 (2018).

\bibitem{ParticleDataGroup:2022pjm}
R.L. Workman et al. (Particle Data Group),
``Review of Particle Physics,''
Prog. Theor. Exp. Phys. 2022, 083C01 (2022).


\bibitem{Oller:2003vf}
J.~A.~Oller,
``The Mixing angle of the lightest scalar nonet,''
Nucl. Phys. A \textbf{727}, 353-369 (2003).

\bibitem{Thiel:2015kuc}
  A.~Thiel {\it et al.} [CBELSA/TAPS Collaboration],
  ``Three-body nature of $N^{\bf *}$ and $\Delta^*$ resonances from sequential decay chains,''
  Phys.\ Rev.\ Lett.\  {\bf 114}, no. 9, 091803 (2015).

\bibitem{CBELSATAPS:2015kka}
V.~Sokhoyan \textit{et al.} [CBELSA/TAPS],
``High-statistics study of the reaction $\gamma p\to p\;2\pi^0$,''
Eur. Phys. J. A \textbf{51} (2015) no.8, 95
[erratum: Eur. Phys. J. A \textbf{51} (2015) no.12, 187].

\bibitem{CBELSATAPS:2022uad}
T.~Seifen \textit{et al.} [CBELSA/TAPS],
``Polarization observables in double neutral pion photoproduction,''
[arXiv:2207.01981 [nucl-ex]].


\bibitem{Burkert:2020inPDG}
see: V. Burkert, E.~Klempt, U.~Thoma, L. Tiator, and R.~L.~Workman,
$\Lambda$ and $\Sigma$ resonances (Fig.82.1), in~\cite{ParticleDataGroup:2022pjm}.

\bibitem{ParticleDataGroup:1986kuw}
M.~Aguilar-Benitez \textit{et al.} [Particle Data Group],
``Review of Particle Properties. Particle Data Group,''
Phys. Lett. B \textbf{170}, 1-350 (1986).

\bibitem{ParticleDataGroup:2012pjm}
J.~Beringer \textit{et al.} [Particle Data Group],
``Review of Particle Physics (RPP),''
Phys. Rev. D \textbf{86}, 010001 (2012).

\bibitem{ParticleDataGroup:2016lqr}
C.~Patrignani \textit{et al.} [Particle Data Group],
``Review of Particle Physics,''
Chin. Phys. C \textbf{40}, no.10, 100001 (2016).

\bibitem{Asner:2016inPDG}
D.~M.~Asner, C. Hanhart, and E. Klempt, ``Resonances", in: \cite{ParticleDataGroup:2016lqr}.

\bibitem{Jacob:1959at}
M.~Jacob and G.~C.~Wick,
``On the General Theory of Collisions for Particles with Spin,''
Annals Phys. \textbf{7}, 404-428 (1959).

\bibitem{Anisovich:2006bc}
A.~V.~Anisovich and A.~V.~Sarantsev,
``Partial decay widths of baryons in the spin-momentum operator expansion method,''
Eur. Phys. J. A \textbf{30}, 427-441 (2006).

\bibitem{VonHippel:1972fg}
F.~Von Hippel and C.~Quigg,
``Centrifugal-barrier effects in resonance partial decay widths, shapes, and production amplitudes,''
Phys. Rev. D \textbf{5}, 624-638 (1972).

\bibitem{Chung:1995dx}
S.~U.~Chung, J.~Brose, R.~Hackmann, E.~Klempt, S.~Spanier and C.~Strassburger,
``Partial wave analysis in K matrix formalism,''
Annalen Phys. \textbf{4}, 404-430 (1995).

\bibitem{Denisenko:2016ugz}
I.~Denisenko {\it et al.},
  ``$N^{\bf *}$ decays to $N\omega$ from new data on $\gamma p\to \omega p$,''
  Phys.\ Lett.\ B {\bf 755}, 97 (2016).

\bibitem{Anisovich:2004zz}
A.~Anisovich {\it et al.},  E.~Klempt, A.~Sarantsev and U.~Thoma,
 ``Partial wave decomposition of pion and photoproduction amplitudes,''
  Eur.\ Phys.\ J.\ A {\bf 24}, 111 (2005).

\bibitem{Anisovich:2007zz}
  A.~V.~Anisovich, V.~V.~Anisovich, E.~Klempt, V.~A.~Nikonov and A.~V.~Sarantsev,
 ``Baryon-baryon and baryon-antibaryon interaction amplitudes in the spin-momentum operator expansion method,''
  Eur.\ Phys.\ J.\ A {\bf 34}, 129 (2007).

\bibitem{Garcia-Martin:2011iqs}
R.~Garcia-Martin, R.~Kaminski, J.~R.~Pelaez, J.~Ruiz de Elvira and F.~J.~Yndurain,
``The Pion-pion scattering amplitude. IV: Improved analysis with once subtracted Roy-like equations up to 1100 MeV,''
Phys. Rev. D \textbf{83},  074004 (2011).

\bibitem{Garcia-Martin:2011nna}
R.~Garcia-Martin, R.~Kaminski, J.~R.~Pelaez and J.~Ruiz de Elvira, ``Precise determination of the $f_0(600)$ and $f_0(980)$ pole
parameters from a dispersive data analysis,'' Phys. Rev. Lett. \textbf{107}, 072001  (2011).

\bibitem{Wang:2022vga}
Z.~Q.~Wang, X.~W.~Kang, J.~A.~Oller and L.~Zhang,
``Analysis on the composite nature of the light scalar mesons $f_0(980)$ and $a0(980)$,''
Phys. Rev. D \textbf{105} no.7, 074016 (2022).

\bibitem{Guo:2012yt}
Z.~H.~Guo, J.~A.~Oller and J.~Ruiz de Elvira,
``Chiral dynamics in form factors, spectral-function sum rules, meson-meson scattering and semi-local duality,''
Phys. Rev. D \textbf{86}, 054006 (2012).

\bibitem{Guo:2012ym}
Z.~H.~Guo, J.~A.~Oller and J.~Ruiz de Elvira,
``Chiral dynamics in U(3) unitary chiral perturbation theory,''
Phys. Lett. B \textbf{712}, 407-412 (2012).

\bibitem{Pelaez:2019eqa}
J.~R.~Pelaez, A.~Rodas and J.~Ruiz De Elvira,
``Global parameterization of $\pi \pi $ scattering up to 2 ${\mathrm {\,GeV}}$,''
Eur. Phys. J. C \textbf{79} no.12, 1008  (2019).

\bibitem{Pelaez:2022qby}
J.~R.~Pelaez, A.~Rodas and J.~Ruiz~de Elvira,
``The $f_0(1370)$ controversy from dispersive meson-meson scattering data analyses,''
[arXiv:2206.14822 [hep-ph]].

\bibitem{Sarantsev:2021ein}
A.~V.~Sarantsev, I.~Denisenko, U.~Thoma and E.~Klempt,
``Scalar isoscalar mesons and the scalar glueball from radiative $J/\psi$ decays,''
Phys. Lett. B \textbf{816}, 136227 (2021).


\bibitem{HADES:2020kce}
J.~Adamczewski-Musch \textit{et al.} [HADES],
``Two-pion production in the second resonance region in ${\pi}^-p$ collisions with the High-Acceptance Di-Electron Spectrometer (HADES),''
Phys. Rev. C \textbf{102}, no.2, 024001 (2020).

\bibitem{Manley:1992yb}
D.~M.~Manley and E.~M.~Saleski,
``Multichannel resonance parametrization of $\pi N$ scattering amplitudes,''
Phys. Rev. D \textbf{45}, 4002-4033 (1992).


\end{thebibliography}
\end{document}